%% file: ms.tex
\documentclass[12pt,preprint]{aastex}
\usepackage{graphicx,lscape}
\def\ea{\it et al. \rm} 
 
\def\ein{{\sl Einstein }} 
 
\def\ros{{\sl ROSAT }} 
\def\cha{{\sl Chandra }} 
\def\xmm{XMM-{\sl{Newton} }} 
\def\axp{{1E2259+586 }}

\shorttitle{Proper Morion of \axp}
\shortauthors{\"{O}gelman \& Tepedelenlio\v{g}lu}
\begin{document}
  
  \title{On the proper motion of AXP 1E2259+586}
\author{H. \"{O}gelman\altaffilmark{1,2}
\email{ogelman@cow.physics.wisc.edu}
 and
E. Tepedelenlio\v{g}lu\altaffilmark{1}}
\email{emre@cow.physics.wisc.edu}
\altaffiltext{1}{Department of Physics, University of
Wisconsin-Madison, 1150 University Ave., Madison, WI 53703, USA}
\altaffiltext{2}{Faculty of Engineering and Natural Sciences, Sabanci
University, Orhanli Tuzla, Istanbul 34956, Turkey}
\begin{abstract} 
  \axp belongs to the group of anomalous X-ray pulsars (AXP) which are
  thought to be magnetars. Accurate measurements of the proper motion
  for these interesting objects have not been made. In this work we
  use the data obtained by satellites {\sl{ROSAT}}, \cha and \xmm
  taken in a period of 12 years to measure the change in position of
  \axp with respect to the three field sources. Our work yielded an
  upper limit on the proper motion of $\sim170$ mas/yr which
  corresponds to a transverse speed of 2500 km~s$^{-1}$, for an
  assumed distance of 3 kpc. We also used this upper limit to put a
  lower limit on the age of the CTB 109 of $\sim6$ kyr, assuming they
  are associated. The consistency of right ascension and declination
  measurements by different satellites suggest that we can average
  these measurements to get the best to date position of \axp.
\end{abstract}

\keywords{Anomalous X-ray Pulsars: \objectname{1E2259+586} --
  AXPs: proper motion -- stars: magnetars}
  
 \section{Introduction} 
  
  Magnetars (AXPs \& SGRs) being neutron stars, must be born during
core-collapse supernovae events. Hence a supernova remnant is expected
to be around. Sure enough 1E2259+586 is associated with CTB109 and AXP
1845-0258 with SNR G29.6+0.1 \citep{woods}. These associations assume
magnetar kick-velocities are similar to those of radio pulsars. It
would be very attractive to measure some magnetar proper motions
directly. This work aims to do that. We have made use of the published
and archival X-ray data to measure the proper motion of 1E2259+586.

In order to eliminate pointing accuracy uncertainties of the
satellites we measured the position of 1E2259+586 relative to other
X-ray point-sources in the same field. In Table~\ref{table1} we list
the archival X-ray observations where we could identify the right
ascension ($\alpha$) and declination ($\delta$) difference between
1E2259+586 and the three reference sources.

\section{Observations}

Since its first discovery Anomalous X-ray Pulsar (AXP) \axp has been
observed with various X-ray missions many times. These observations
started as early as \ein and continued all the way to state-of-the-art
satellites like \cha and XMM-{\sl{Newton}}. For this work we choose
the observations that satisfy the following criteria; 1) Data taken by
detectors that have high spatial resolution ($\sim 10''$), 2) All
three reference sources (see Fig.~\ref{fig1}) are in the field of
view, 3) Exposure is long or the detector is sensitive enough so that
all three reference sources are detected. In Table~\ref{table1} we
list the observations and the name of their respective observatories
that we have used to get the positions of three selected sources
together with the position of 1E2259+586.

In order to get the positions of the selected sources in the field of
view of the pulsar we extracted unbinned images of all five
observations. For data sets obtained with \ros PSPC and HRI, and \cha
ACIS we used the CIAO (version 3.2) tool
WAVDETECT{\footnote{http://asc.harvard.edu/ciao/threads/wavdetect/}}
to detect all sources in the field of view. For \xmm observation we
used the SAS (version 6.1.0) tool EBOXDETECT that performs a sliding
box detection, using locally estimated background (see SAS data
analysis threads web
page{\footnote{http://xmm.vilspa.esa.es/external/xmm\_sw\_cal/sas\_frame.shtml}}).
Although using WAVDETECT on \ros HRI will not yield accurate Point
Spread Function (PSF) sizes, this should not effect our results
because we are only interested in the central position of the sources.

\section{Determining the Proper Motion of \axp}

In Table~\ref{table1} we show the measured coordinates of all three
reference sources together with the X-ray pulsar. Each of these
measurements have been done for five different times. For each
observation we calculated the distances of the reference sources to
the pulsar in the direction of right ascension ($\alpha$) and
declination ($\delta$). The individual errors were calculated from the
positional accuracies of the respective instruments. PSPC detector on
board ROSAT has a spatial on-axis resolution of 25 arcsec FWHM at 0.93
keV. The other detector HRI, however, has relatively high spatial
resolution of $\sim 2''$.  Later, more advanced instruments such as
\cha ACIS-S and \xmm MOS have also good angular resolution. For an
on-axis source ACIS-S point-spread function (PSF) is under sampled by
the 0\farcs492$\times$0\farcs492 CCD pixels. On the other hand the
EPIC-MOS camera has a spatial resolution of $5''$ FWHM, which is
limited by the mirrors. Since all four sources have different spectra
we chose these typical limiting resolutions as the error radii. These
errors were then propagated with standard error propagation to get the
uncertainty on each angular separation. To each of these six data sets
we made a error weighted least-squares fit with a straight line,
$y=ax+b$ (see Figure~\ref{fit}).

Through the described method, for each data set we found the best fit
value. The parameter $a$ (slope) was varied around its best fit value
(18 mas/yr) while $b$ was kept constant at the value determined by the
best fit. For each different value of $a$ a $\chi^{2}$ value was
calculated. This distribution is approximately a parabola. To
calculate the combined $\chi^{2}$ distribution for all three data sets
that give the proper motion for $\alpha$ (or $\delta$) we then added
the individual $\chi^{2}$s and found the minima to get a best fit
value (see Figure~\ref{chisq}). To get the errors associated with the
proper motions calculated we used the method described by
\citet{lampton}. $1\sigma$ error associated with the proper motion
were found by adding 10.7\footnote{This is the value of the reduced
chi-square ($\chi^{2}/\nu$) corresponding to the probability of
observing a value of chi-square larger than $\chi^{2}$ with $\nu$
degrees of freedom. }  to our minimum $\chi^{2}$ for total of 9
degrees of freedom and found the width of the curve at that point
(Figure~\ref{chisq}). We give the values of the best fit and it's
$1\sigma$ error in Table~\ref{table2}.

\section {Absolute position of \axp} Since the data set on Table~\ref{table1}
represents also a summary of all the X-ray position measurements of
\axp, the consistency of the right ascension, declination measurements
by different satellites suggest that we can average these measurements
to get the best to date position of \axp. The variance in the data set
would also yield the error in this average position measurement. The
results in (J2000) are, $\alpha= 23^{h}01^{m}08\fs 21 \pm 2\farcs2$
and $\delta= +58\arcdeg52\arcmin44\farcs8 \pm 2\farcs2$. These
coordinates, within the error-bars include candidates from previous
searches for the optical counterpart (\citet{hulleman},
\citet{hulleman00}). Optical identifications, or upper limits are
crucial for AXPs since it determines the mass and nature of the donor
component, hence the energy emission mechanism.

\section {Discussion} 

The uncertainties calculated being larger than the proper motion
indicates that the available data is not sufficient to determine a
definite proper motion. However, we can put an upper limit on it. The
maximum angular speed \axp can be calculated by the simple formula
$(\mu^{2}_{\alpha}+\mu^{2}_{\delta})^{1/2}$ (see Table~\ref{table2})
which gives us an upper limit on the proper motion of $\sim 170$
mas/yr. This together with the adopted distance of 3 kpc
\citep{kothes} gives us a maximum transverse velocity of 2500
km~s$^{-1}$. Although from the best fit values we can calculate a
direction for the proper motion, the uncertainty being too high does
not make the effort worth while.

\begin{figure}
\plottwo{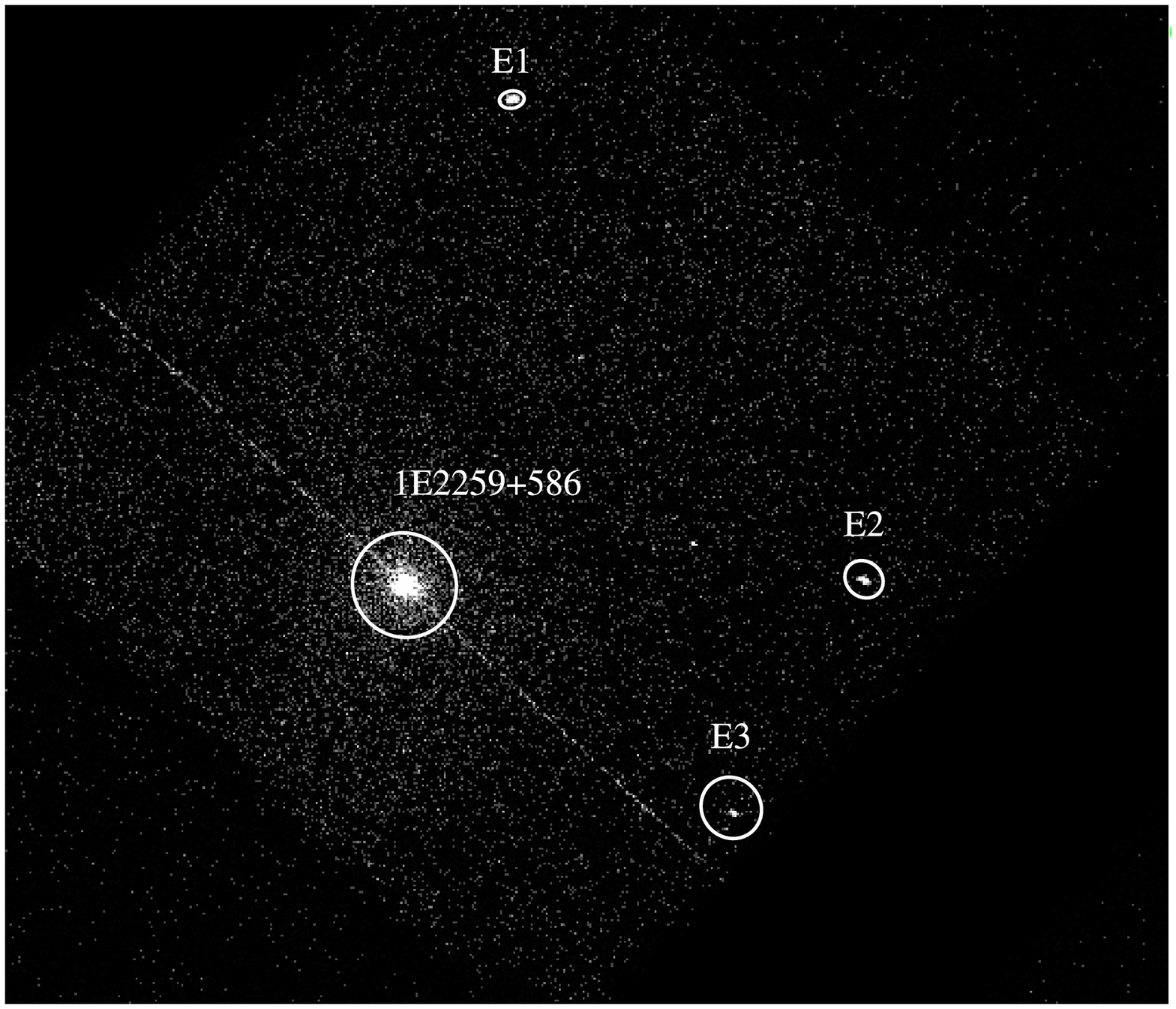}{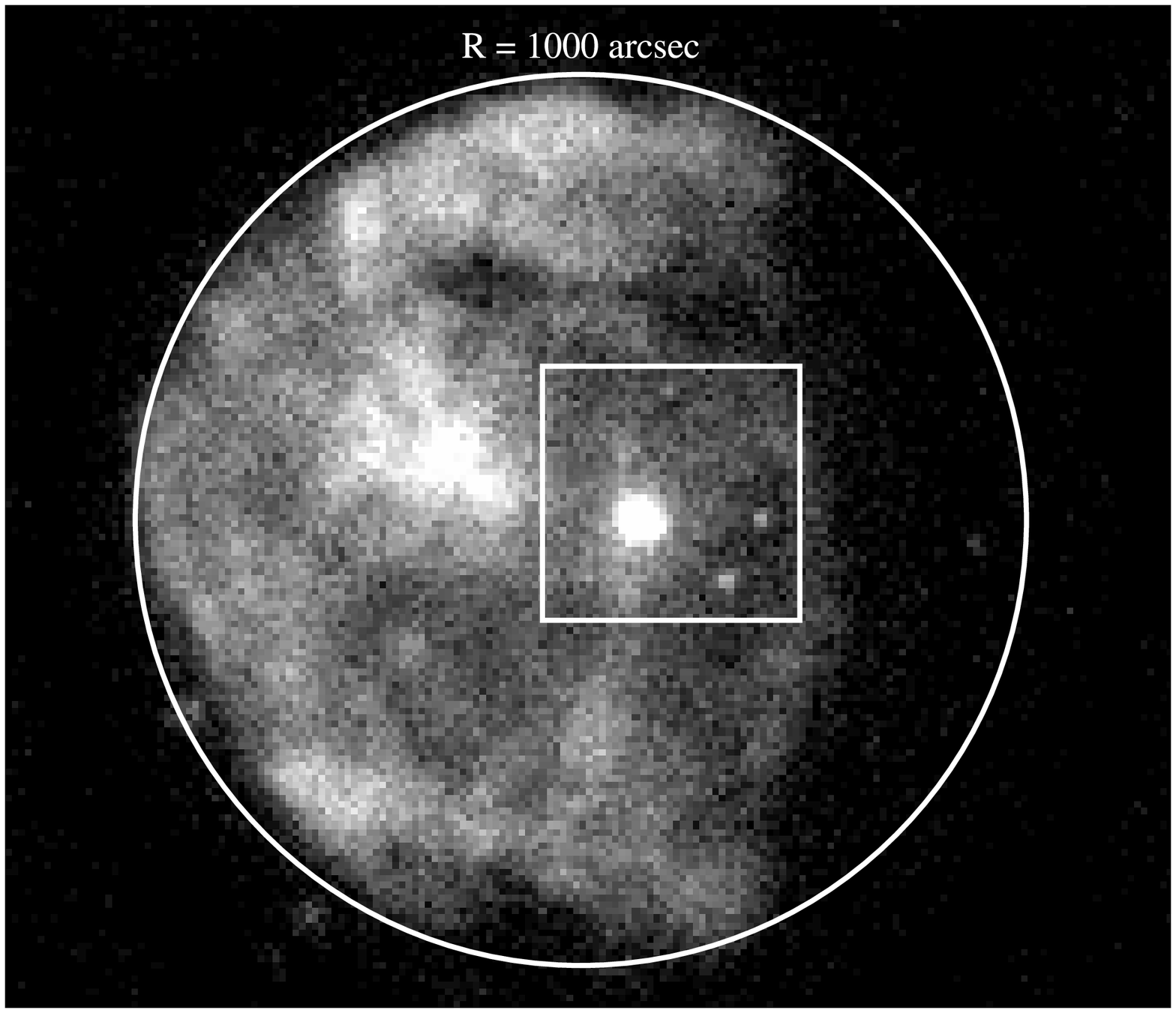}
\caption{\small{({\it{Left}})The region that is blown up here is shown
with a box on the right hand side image. The sources around 1E2259+586
and their respective names adopted in this paper. The image is taken
from \cha observation, see Table 1. ({\it{Right}}) The circle
approximately cocentered with SNR. The radius of the circle is
$1000''$, which is a rough estimate of the size of the remnant.}}
\label{fig1}
\end{figure}

\begin{figure}
\plottwo{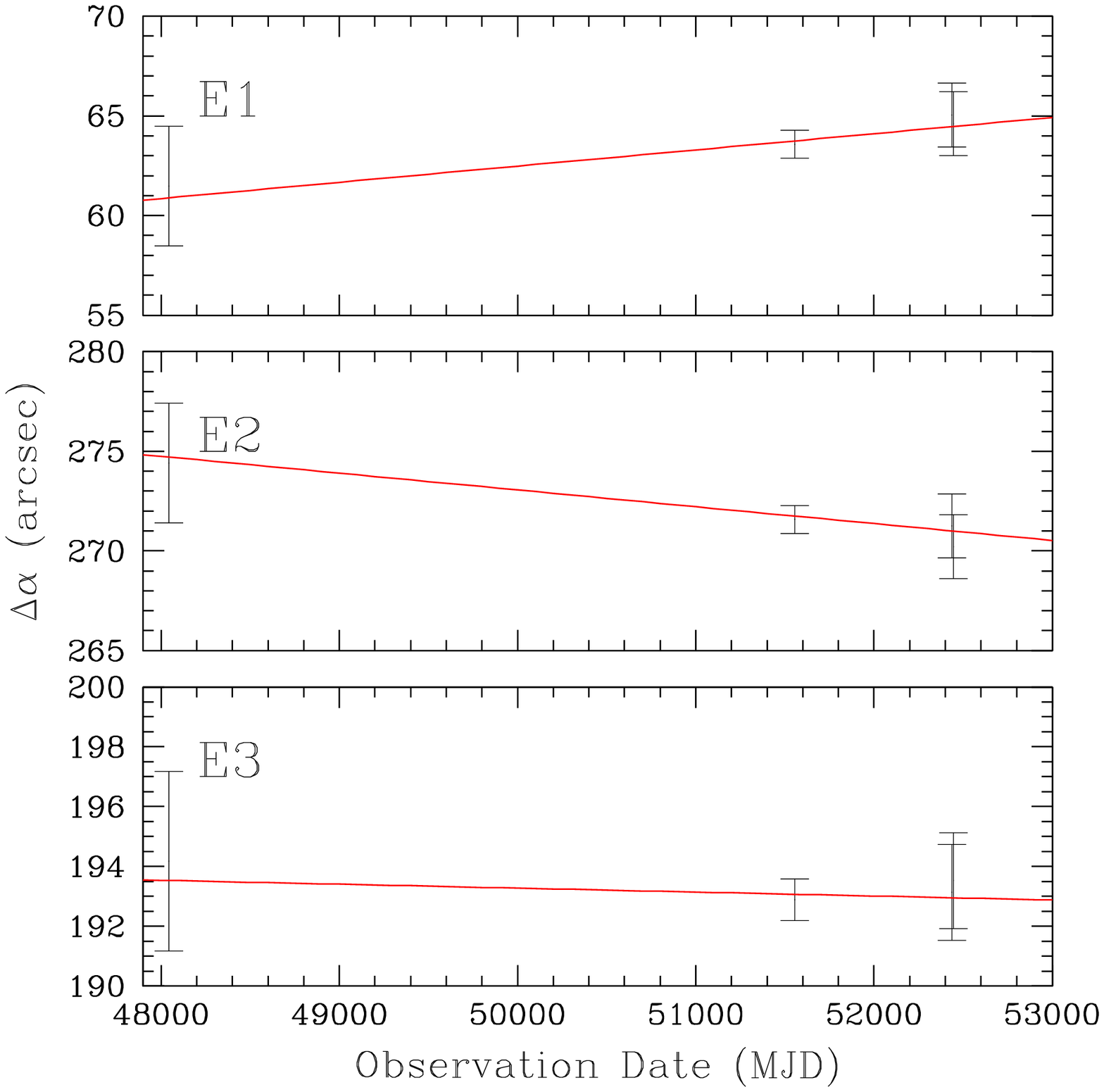}{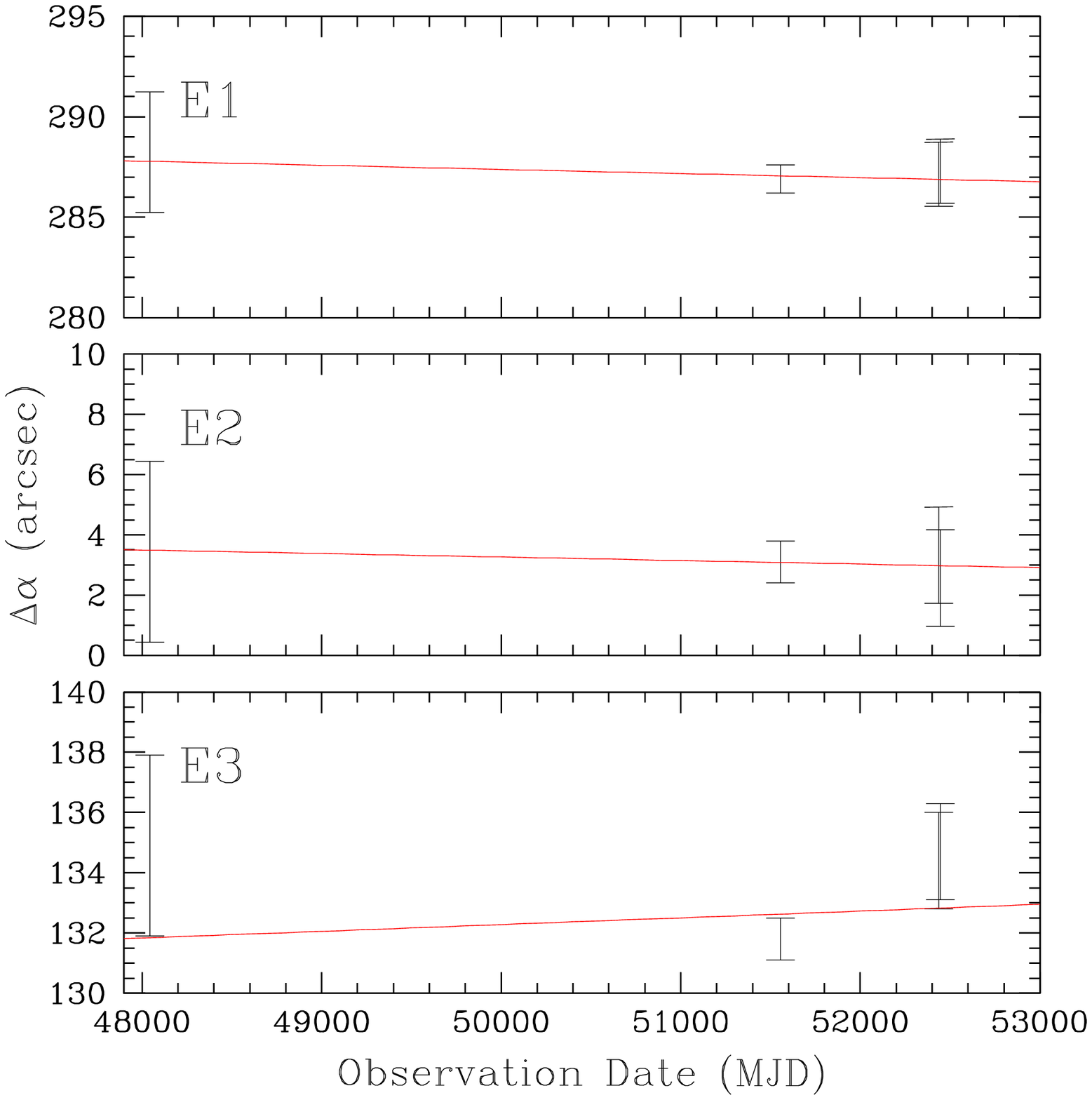}
\caption{\small{The angular distance between \axp and the respective
X-ray source in the direction of $\alpha$ ({\it{Left}}) and $\delta$
({\it{Right}}) versus the date of the observation. The best linear fit
to data is shown with a straight line in each case. We have taken out
the datum for the ROSAT PSPC observation just for plotting
purposes. It was, however, included in the fit.}}
\label{fit}
\end{figure}

\begin{figure}
\plottwo{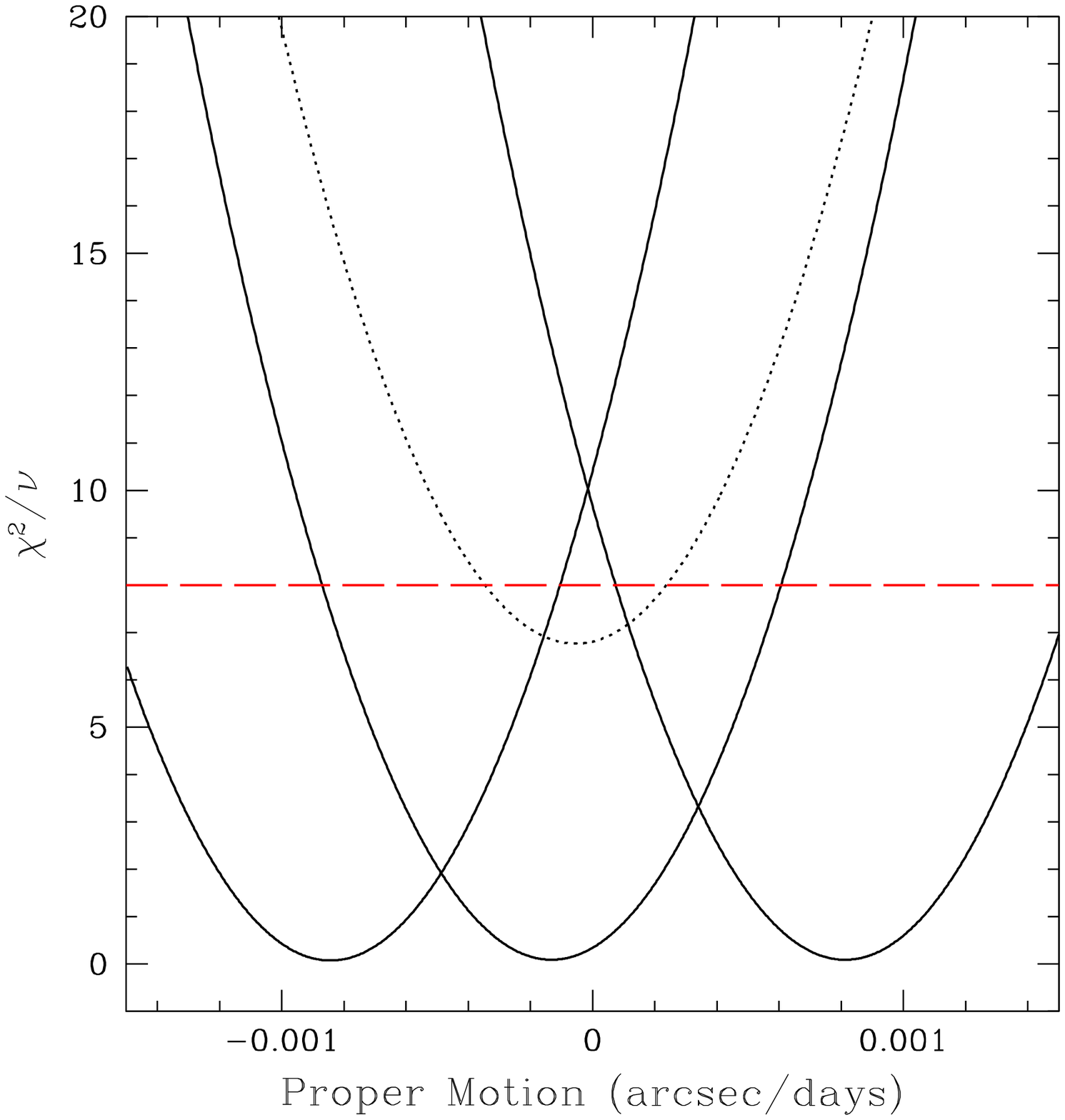}{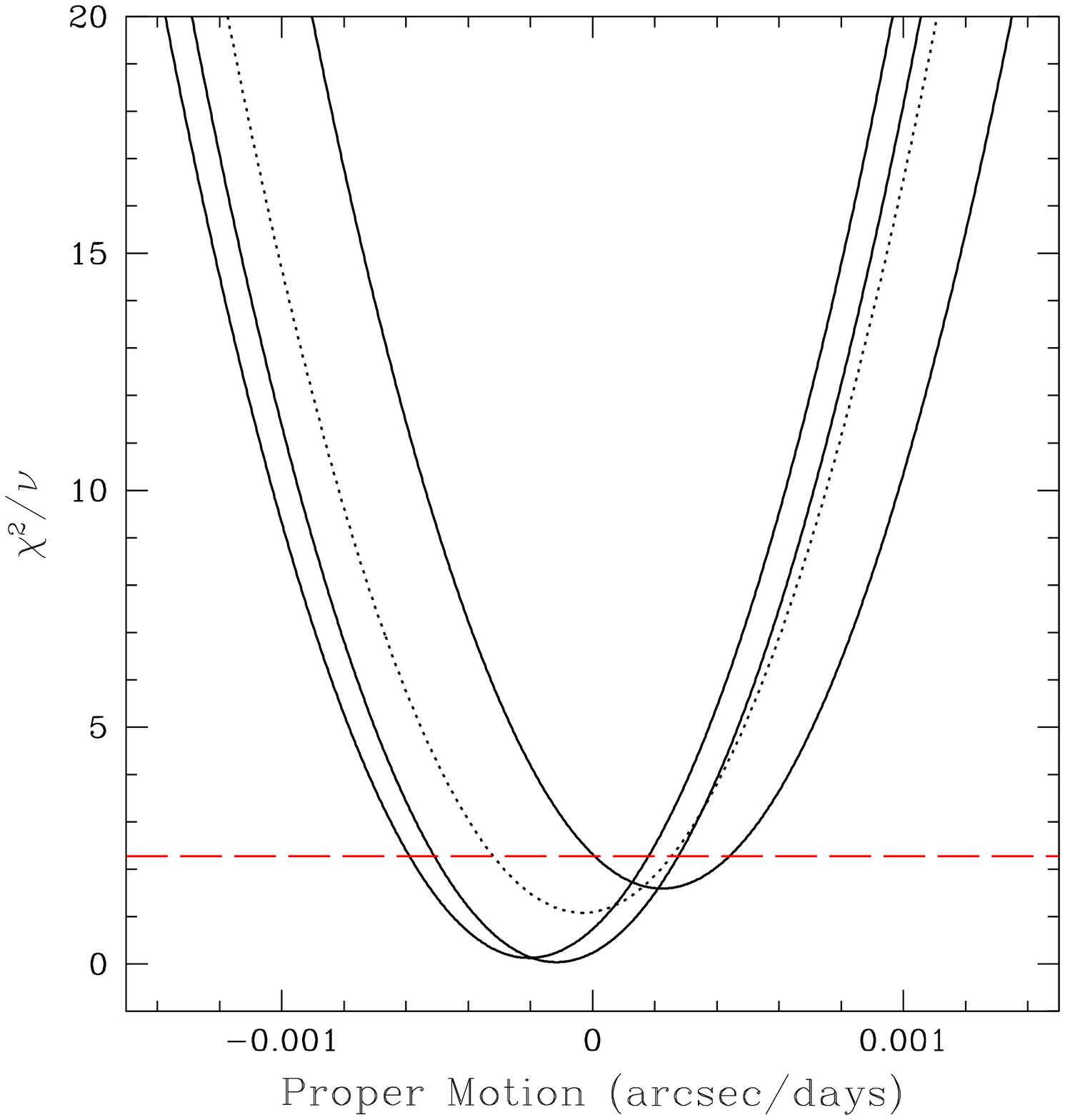}
\caption{\small{$\chi^{2}$ per degrees of freedom ($\nu$) versus proper
motion in the direction of $\alpha$ ({\it{Left}}) and $\delta$
({\it{Right}}). Solid curves are the $\chi^{2}$ distribution of the
individual fits whereas the dotted line is the distribution of all
three combined (average). $\nu$ is 3 for individual fits and 9 for the
combined distribution. The dashed horizontal line is the marker for
the $1\sigma$ error level. See text for the explanation.}}
\label{chisq}
\end{figure}

Associations, in general, between neutron stars and SNRs are usually
judged on criteria such as agreement in age/distance, positional
coincidence and evidence from proper motion. Distance estimates for
AXPs have relatively big uncertainties, and there is no evidence that
their characteristic ages are reliable age estimators. For \axp we
have two different age estimators; the age of the host remnant
($\sim$20 kyr, \citet{rho}) and also the spin-down age
($\sim$220 kyr). The age derived from the SNR assumes that they
are physically associated. This method is a rough estimate for the age
of the SNR and depends on the assumed spectral model for the X-ray
emitting gas that surrounds the remnant. Our approach gives a lower
limit for the age of 1E2259+586. Although we also assume that the SNR
and the pulsar are associated, we make no assumptions on the spectral
properties of CTB109 and its distance. The SNR is roughly spherical in
shape and \axp is close to its geometric center. We estimated the
radius of this remnant as $\sim 1000''$ (see Figure~\ref{fig1}). This
radius together with the measured upper limit for the proper motion
gives us a lower limit on the age of the pulsar of $\sim6$ kyr. This
limit is consistent with all other age estimates and is not very
restrictive. However it is completely independent of all other methods
used to determine the age. This limit is also very close to the age
$\sim9$ kyr determined by \citet{sasaki} from several \xmm
observations. 

The upper limit of 2500 km~s$^{-1}$ for the proper motion is
consistent with that of those expected for radio
pulsars. \citet{arzoumanian} calculated the velocity distribution of
radio pulsars. They find that $\sim$15\% of pulsars have velocities
greater than 1000 km~s$^{-1}$ and the distribution function
asymptotically approaches zero beyond 700 km~s$^{-1}$ for a two
component distribution. Although our upper limit is higher than all
(but PSR B1552-31, \citep{manchester}) known pulsar velocities it is
right on the upper limit of those that are expected for radio pulsars.

Furthermore, birth properties of magnetars need not be the same as
radio pulsars. For example, the magnetar model predicts that soft
gamma repeaters may be born with velocities $\ga1000$ km~s$^{-1}$
(e.g. see \citet{thompson00}). We should also note that the velocity
calculated here is a function of the distance, which is not well known
and has an uncertainty $\ga 50\%$.

Another way of obtaining an estimate for the speed is to check the
potential consequences high transversal velocities may cause. For
instance, the speed of sound in the medium surrounding the pulsar is
on the order of $\ga$5000 km~s$^{-1}$ assuming a gas temperature of 1
MK and mean particle density of 1 cm$^{-3}$. If \axp had a proper
motion larger than this than we would expect a shock front to form, no
such thing has been detected. Also, the previous works that measured
the absolute position of the pulsar for two different observations
were not conclusive. The current best position \citep{patel} when
compared to the previous best position \citep{hulleman00} for \axp
yields no discrepancy within the uncertainties. Hence we conclude that
the proper motion measurements that has been made in this work are the
best so far. There are cases where optical sources are obscured by
galactic dust absorbtion. Hard X-rays may be the only handle we have
on the proper motion of some of these objects. However, relative
proper motion measurements are usually done in optical and
radio-astronomy since the requirements are: 1) there be enough sources
in the field, 2) the accuracy of position measurements are high. The
current observatories allows us to meet both of these criteria. Thus
confirming the coming of age in X-ray astronomy.

\clearpage
\clearpage
\input{table1.tex}
\clearpage
\input{table2.tex}
\clearpage
\clearpage

\end{document}

%% file: table1.tex
\begin{landscape}
\begin{table}
\scriptsize
\caption{RA and DEC of selected sources around the pulsar, for
different observations\label{table1}}
\begin{tabular}{llc||lc||lc||lc||lc}
\tableline
Mission & \multicolumn{2}{c}{\ros} & \multicolumn{2}{c}{\ros}& \multicolumn{2}{c}{\cha}& 
\multicolumn{2}{c}{\xmm}& \multicolumn{2}{c}{\xmm} \\
Detector &\multicolumn{2}{c}{PSPC}&\multicolumn{2}{c}{HRI}&\multicolumn{2}{c}{ACIS}& 
\multicolumn{2}{c}{MOS}& \multicolumn{2}{c}{MOS} \\
Date&\multicolumn{2}{c}{1991-07-09}&\multicolumn{2}{c}{1992-01-08}&\multicolumn{2}{c}{2000-01-12}&\multicolumn{2}{c}{2002-06-11}&\multicolumn{2}{c}{2002-06-21}\\
Duration 
(ks)&\multicolumn{2}{c}{33.37}&\multicolumn{2}{c}{22.41}&\multicolumn{2}{c}{19.13}&\multicolumn{2}{c}{52.53}&\multicolumn{2}{c}{31.05}\\
\tableline\tableline
\multicolumn{11}{c}{Sources and their respective coordinates, ($\alpha$,$\delta$), for 
each observation}\\
1E2259+586  & 345.2845 & +58.8783 & 345.2828 & +58.8805 & 345.2845 & +58.8792 & 345.2842 
& +58.8788 & 345.2850 & +58.8788\\
E1 & 345.2491 & +58.9628 & 345.2498 & +58.9606 & 345.2504 & +58.9589 & 345.2492 & 
+58.9585 &345.2502& +58.9586\\
E2 & 345.1378 & +58.8791 & 345.1353 & +58.8815 & 345.1386 & +58.8801 & 345.1384 & 
+58.8797 &345.1397 & +58.8795\\
E3 & 345.1785 & +58.8416 & 345.1784 & +58.8430 & 345.1809 & +58.8426 & 345.1804 & 
+58.8415 &345.1809 & +58.8414\\
\tableline
\end{tabular}
\end{table}
\end{landscape}

%% file: table2.tex
\begin{table}
\begin{center}
\caption{Statistics for the simultaneous fits done for $\Delta\alpha$
and $\Delta\delta$ for each selected source, see Figure~\ref{fit}. We
used a linear function of the form $y=ax+b$. The errors given are
1$\sigma$.\label{table2}}
\begin{tabular}{lcccc}
\tableline
   & $\chi^{2}$/dof & $a$ & $\Delta a$ &Proper Motion ($\mu$)\\
   &           & (arcsec/day) & (arcsec/day)& (mas/yr)\\
\tableline\tableline
$\alpha$ &61.2/9&-0.000054&0.000285&$18\pm 104$\\
$\delta$ &9.9/9 &-0.000033&0.000285&$12\pm 104$\\
\tableline
\end{tabular}
\end{center}
\end{table}